\documentclass[aps,prb,twocolumn,floatfix]{revtex4-1}

\usepackage[utf8]{inputenc}
\usepackage{amsfonts}
\usepackage{amsmath}
\usepackage{amssymb}
\usepackage{bbm}
\usepackage{amsthm}
\usepackage{empheq}
\usepackage{enumitem}
\usepackage{verbatim}
\usepackage[dvips]{graphicx}
\usepackage[dvips]{color}
\usepackage{subfigure}
\usepackage[breaklinks]{hyperref}

\begin{document}

\title{Universal spin wave damping in magnetic Weyl semimetals}
\author{Predrag Nikoli\'c$^{1,2}$}
\affiliation{$^1$Department of Physics and Astronomy, George Mason University, Fairfax, VA 22030, USA}
\affiliation{$^2$Institute for Quantum Matter at Johns Hopkins University, Baltimore, MD 21218, USA}
\date{\today}

\begin{abstract}

We analyze the decay of spin waves into Stoner excitations in magnetic Weyl semimetals. The lifetime of a mode is found to have a universal dependence on its frequency and momentum, and on a few parameters that characterize the relativistic Weyl spectrum. At the same time, Gilbert damping by Weyl electrons is absent. The decay rate of spin waves is calculated perturbatively using the s-d model of itinerant Weyl or Dirac electrons coupled to local moments. We show that many details of the Weyl spectrum, such as the momentum-space locations, dispersions and sizes of the Weyl Fermi pockets, can be deduced indirectly by probing the spin waves of local moments using inelastic neutron scattering.

\end{abstract}

\maketitle

\section{Introduction}\label{intro}

Weyl semimetals are condensed matter realizations of massless fermions with a chiral relativistic three-dimensional spectrum \cite{Ari2010, Burkov2011a, Ran2011}. Topologically protected gapless Fermi ``arc'' states on the system boundaries, and unconventional transport properties such as the intrinsic anomalous Hall effect, set Weyl semimetals apart from other weakly interacting conductors. One way to obtain a Weyl spectrum involves breaking the time-reversal symmetry in a material that has Dirac quasiparticles. The presence of magnetization, for example, will remove the spin degeneracy of a Dirac node by splitting it into a dipole of opposite-chirality Weyl nodes in momentum space. Magnetism then becomes intimately related to the presence of Weyl electrons. Alternatively, Weyl spectrum of itinerant electrons can be created by a broken inversion symmetry, e.g. due to the crystal structure, and then coupled to magnetism if the material possesses additional local moments or undergoes a spin density wave instability. Some of these theoretical scenarios are slowly finding their actualization in experimentally studied magnetic Weyl semimetals \cite{Yang2017, Kuroda2017, Ghimire2018, Sakai2018, Kimata2019, Broholm2020, Zaliznyak2020, Luo2020, Markou2021, Gaudet2021}.

Here we analyze an important imprint of Weyl electrons on the magnetic dynamics -- the damping of spin waves via particle-hole (Stoner) excitations. This basic interaction effect reveals the defining features of the Weyl spectrum, relativity and chirality. We will show that the lifetime of spin waves exhibits a universal dependence on the mode frequency and momentum which can be used to extract detailed properties of the underlying Weyl electrons. By measuring the mode lifetime throughout the first Brillouin zone, it is possible to discern the locations of the Weyl nodes in momentum space, their relative chiralities, slope of the energy versus momentum dispersion, and the size of the Fermi pockets on the Weyl nodes. The spin wave lifetime is obtained from the width of the scattering intensity peaks in inelastic neutron scattering experiments, provided that a sufficient energy resolution is available and other sources of decoherence (thermal broadening, disorder, phonons) do not mask the electronic source.

Even though neutron scattering is a powerful Green's function probe, its ability to detect fermionic quasiparticles is normally ruined by the incoherent continuum of excitations that can absorb an angular momentum quantum. Interestingly, this problem is reduced in Weyl semimetals \cite{Turner2020}, and fortunately it is also possible to indirectly characterize the quasiparticles via collective excitations. The latter has been achieved in the neutron studies of samarium hexaboride (SmB$_6$) \cite{Fuhrman2014, Nikolic2014c}, where the measured dispersion of a ``spin exciton'' has revealed a non-trivial topology of the underlying electronic quasiparticles. An energy gap protects the exciton's coherence in SmB$_6$, but the gapless quasiparticles in Weyl semimetals will generally induce ubiquitous damping of collective modes. Such a damping can in fact reveal the existence and properties of chiral fermionic quasiparticles. The Weyl electron characterization through damping could potentially overcome various issues that plague other approaches, such as correlation effects in the case of band-structure calculations, limited resolution in the case of ARPES, sensitivity to conventional bands (that coexist with Weyl nodes) in transport measurements, etc.

Closely related to the physics we pursue here is the extensively studied damping in metallic ferromagnets \cite{Kambersky1976, Berger1996, Simanek2003, Mills2003, Zhang2004, Tserkovnyak2005, Antropov2006, Kambersky2007, Hickey2009, Tatara2010, Umerski2014, Yudin2018, Batista2018}. Stoner excitations provide a mechanism for the decay of spin waves, and also typically give rise to Gilbert damping \cite{Gilbert2004} -- the dissipated precession of uniform magnetization in an external magnetic field. Many works have been devoted to the calculation of Gilbert damping since it is possible to measure it by ferromagnetic resonance \cite{Bailey2007, Woltersdorf2009} and time-resolved magneto-optical Kerr effect \cite{Iihama2014, Capua2015}. A careful consideration of the relativistic electron dynamics has revealed that Gilbert damping originates in the spin-orbit coupling and depends on the electrons' mass \cite{Hickey2009}. In the case of massless Weyl electrons, we show here that Gilbert damping is absent. However, spin waves unavoidably decay via Stoner excitations \cite{Fisher1971, Prange1972, Cade1977, Isoda1990, Bunemann2001, Costa2003, Tserkovnyak2008}, and their damping features ``non-reciprocity'' -- different polarization modes that carry the same momentum have different damping rates. This accompanies the non-dissipative aspects of chiral spin-momentum locking \cite{Nikolic2019b, Nikolic2020a}. Spin wave ``non-reciprocity'' has been anticipated in spiral magnets \cite{Baryakhtar2019}, magnetic interfaces with a Dzyaloshinskii-Moriya interaction derived from the Rashba spin-orbit coupling \cite{Costa2010, Kim2012d, Stiles2013, Manchon2014, Zakeri2014, Kim2015, Akosa2016}, and observed in several experiments \cite{Yang2015, Tokura2016, Tokura2016b, Matan2017, Boni2018, Weber2018}. In the context of magnetic Weyl semimetals, initial theoretical studies have been focused on the domain wall dynamics \cite{Nomura2019, Araki2020}.

The rest of this paper is organized as follows. Section \ref{secSummary} presents the approach and the main results of the analysis, focusing on the observable physical characteristics of the spin wave damping by Weyl electrons. Section \ref{secDissipation} is devoted to the technical development of the damping theory. It contains separate derivations of the dissipative terms in the effective spin action (\ref{secEffLagrangian}), spin wave damping (\ref{secSpinWaves}), and Gilbert damping from the semiclassical field equation (\ref{secFieldEq}). The last section \ref{secConclusions} summarizes the conclusions and discusses the broader applicability and limitations of the damping theory.

\section{Summary of the results}\label{secSummary}

\begin{figure}
\subfigure[{}]{\includegraphics[width=3.0in]{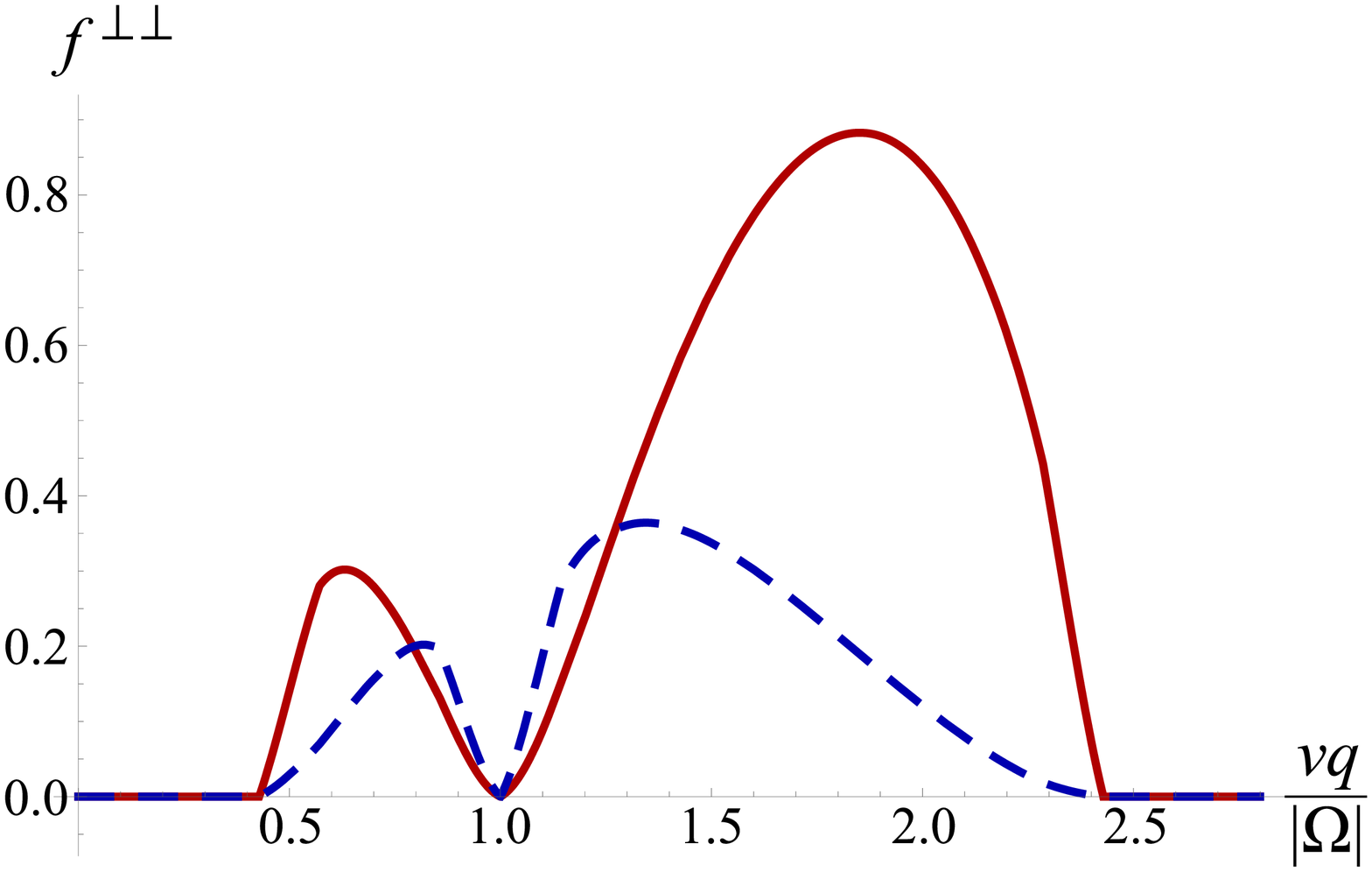}}
\subfigure[{}]{\includegraphics[width=3.0in]{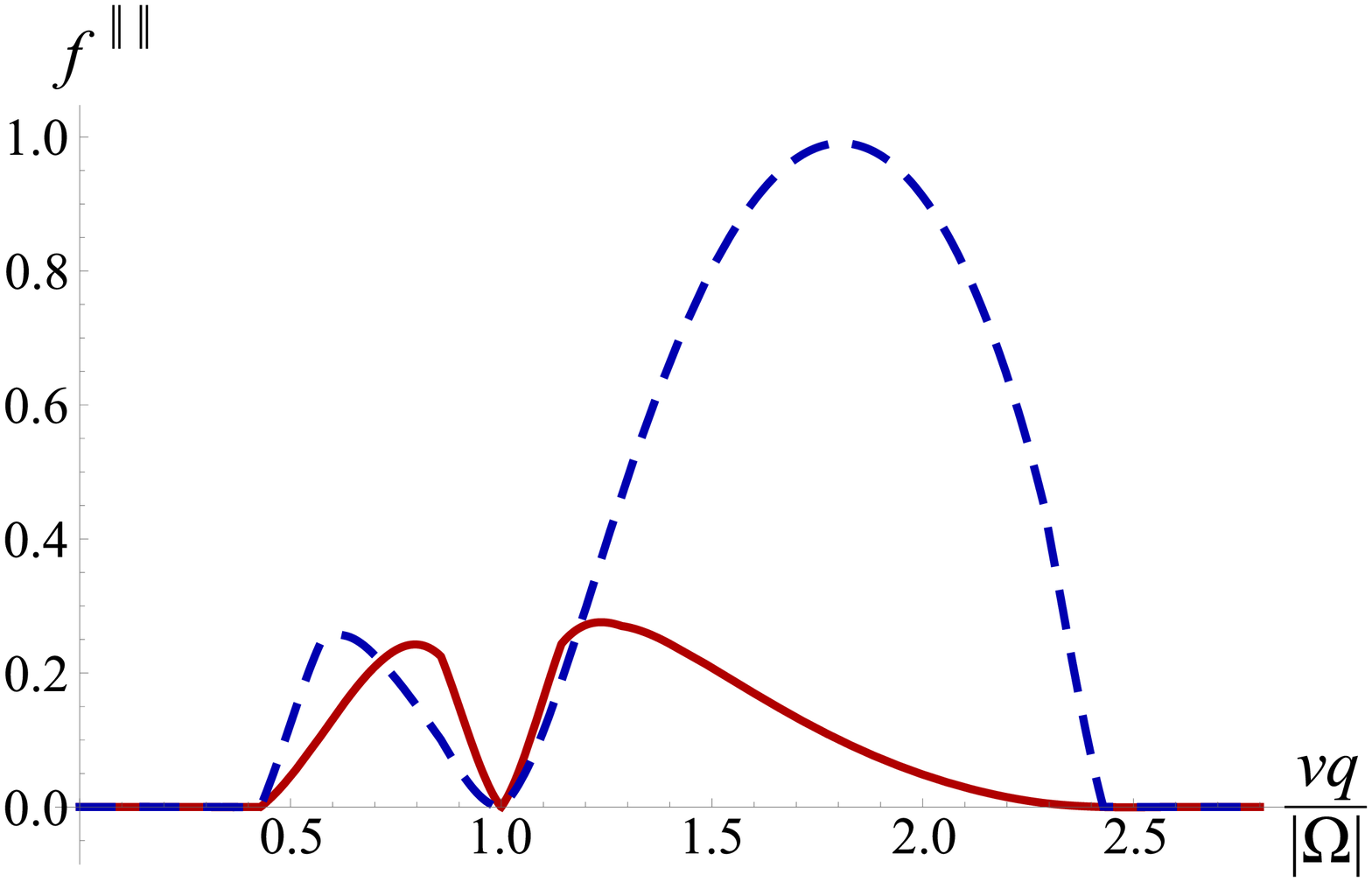}}
\caption{\label{damping-tl}The plots of functions (a) $f^{\perp\perp}$ and (b) $f^{\parallel\parallel}$ for the damping rates of transverse and longitudinal spin waves respectively, contributed by the Fermi surfaces on a particular pair of Weyl nodes. Solid red lines are for the same-chirality nodes, and the dashed blue lines are for the opposite-chirality nodes. $|\Omega|=1.4|\mu|$ was assumed in this example.}
\end{figure}

\begin{figure}
\includegraphics[width=3.3in]{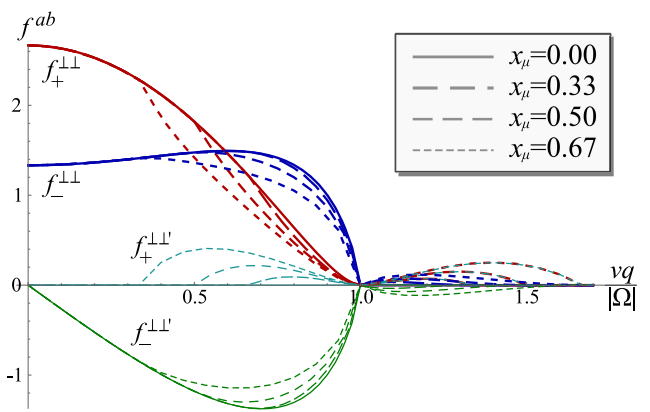}
\caption{\label{damping-all}The plots of selected universal functions $f^{ab}$ featured in the damping rate $\gamma \sim \Omega^2\, f(vq/|\Omega|;x_\mu)$. The functions are parametrized by $x_\mu = 2|\mu/\Omega|$, with finer dashes corresponding to larger Weyl Fermi pockets (solid lines refer to the Fermi level that crosses the Weyl nodes). Shown functions include transverse ($\perp\perp$) and chiral ($\perp\perp'$) damping channels shaped by electron scattering between equal-chirality ($+$) and opposite-chirality ($-$) Weyl nodes. Longitudinal channels ($\parallel\parallel$) are similar to the shown transverse channels, compare with Fig.\ref{damping-tl}.}
\end{figure}

In this paper, we work with the s-d model of Weyl electrons coupled to local moments. We perturbatively calculate the dissipative non-Hermitian parts of the moments' effective action, which determine the rate $\gamma$ of spin wave damping. $\gamma$ also depends on the magnetic order and the wave's propagation direction relative to the magnetization, but it is always controlled by the components of the universal damping rate tensor given by
\begin{equation}\label{DampingRate2}
\gamma^{ab}_{mn}(q) = \frac{a^3 J_{\textrm{K}}^{2} \Omega^2}{128\pi Sv^3}\; f_{mn}^{ab}\left(\frac{|\Omega|}{vq},\frac{|\Omega|}{2|\mu|};\textrm{sign}(\mu,\Omega)\right)
\end{equation}
for ferromagnetic local moments of spin magnitude $S$. The upper indices $a,b\in\lbrace x,y,z\rbrace$ refer to spin projections. The universal scaling functions $f_{mn}^{ab}$ are dimensionless, the factor $a^3$ is the unit-cell volume of the local moment's lattice, $J_{\textrm{K}}$ is the Kondo or Hund coupling energy scale, $v$ and $\mu$ are the Fermi velocity and Fermi energy of the Weyl electrons respectively, and $\Omega$ is the real spin wave frequency (we use the units $\hbar=1$). The spin wave momentum ${\bf q}$ in this expression is measured relative to the difference $\Delta{\bf Q}={\bf Q}_m-{\bf Q}_n$ between the wavevectors ${\bf Q}_m, {\bf Q}_n$ of any two Weyl nodes in the first Brillouin zone. Coherent collective excitations that span the entire first Brillouin zone can be used to separately address many pairs of Weyl nodes -- by tuning the total wavevector $\Delta{\bf Q}+{\bf q}$ to the vicinity of $\Delta{\bf Q}$. Representative functions $f_{mn}^{ab}$ for the Weyl nodes with finite Fermi surfaces are plotted in Figures \ref{damping-tl} and \ref{damping-all}

We make analytical progress and gain valuable physical insight through several idealizations: all Weyl nodes are assumed to be identical, spherically symmetric and living at the same node energy. Their chiralities $\chi_m=\pm1$ and locations ${\bf Q}_m$ are arbitrary (as long as the total chirality in the first Brillouin zone vanishes). Under these conditions, only three tensor components of $\gamma^{ab}$ are finite and independent, $\gamma^{\parallel\parallel}$, $\gamma^{\perp\perp}$ and $\gamma^{\perp\perp'}$. Here and throughout the paper $\parallel$ indicates the spin direction parallel to the mode's wavevector ${\bf q}$, and $\perp,\perp'$ are the spin directions which are perpendicular to ${\bf q}$ and each other. The full expression for damping rates is presented in Section \ref{secSpinWaves}; in Weyl ferromagnets, it becomes
\begin{equation}\label{DampingRate1}
\gamma_{mn}^{\phantom{x}} = \gamma_{mn}^{\perp\perp} \pm \gamma_{mn}^{\perp\perp'}
\end{equation}
for the two polarizations of spin waves propagating along the magnetization direction.

\begin{figure}[!t]
\subfigure[{}]{\includegraphics[width=1.68in]{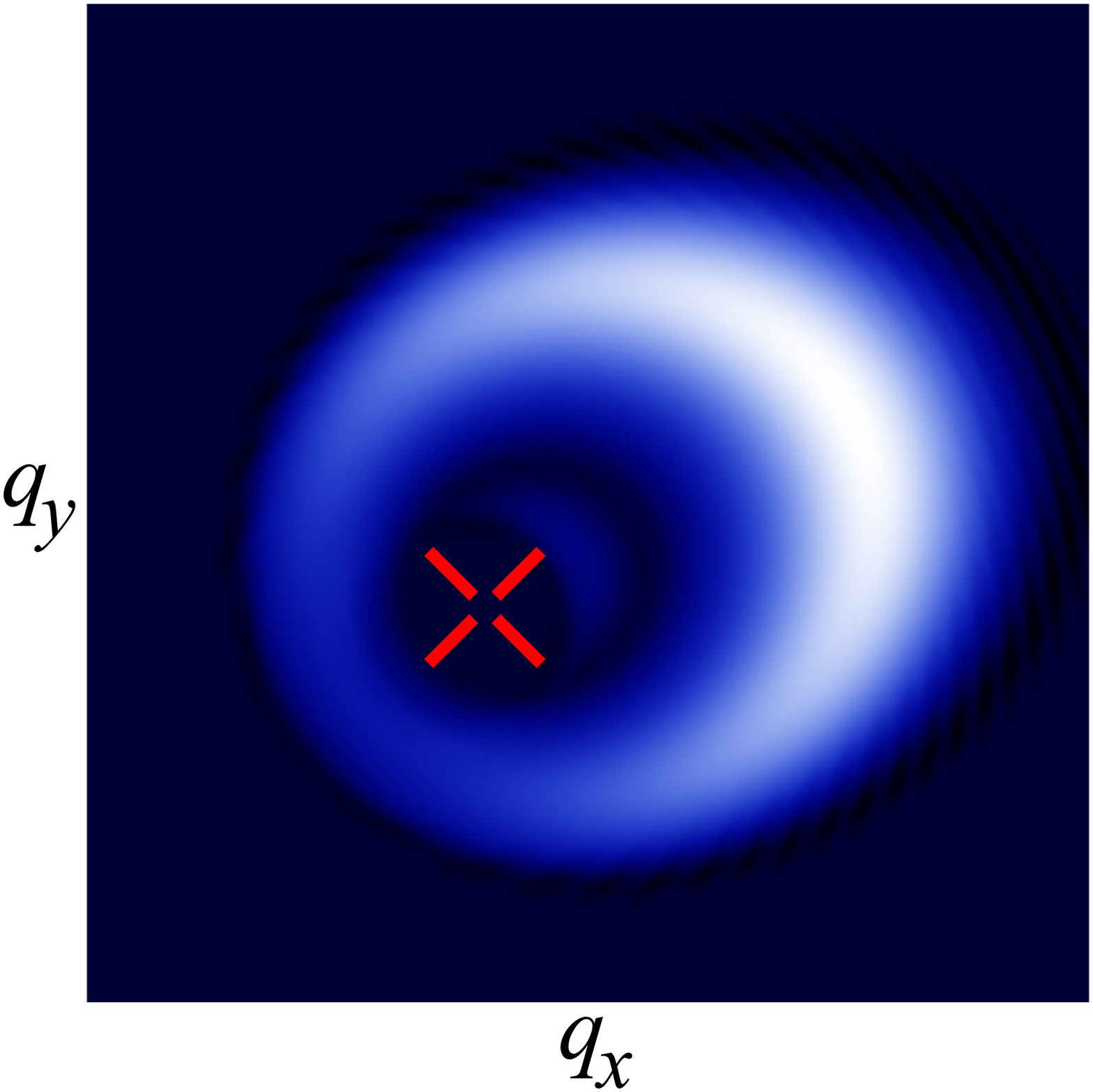}}
\subfigure[{}]{\includegraphics[width=1.68in]{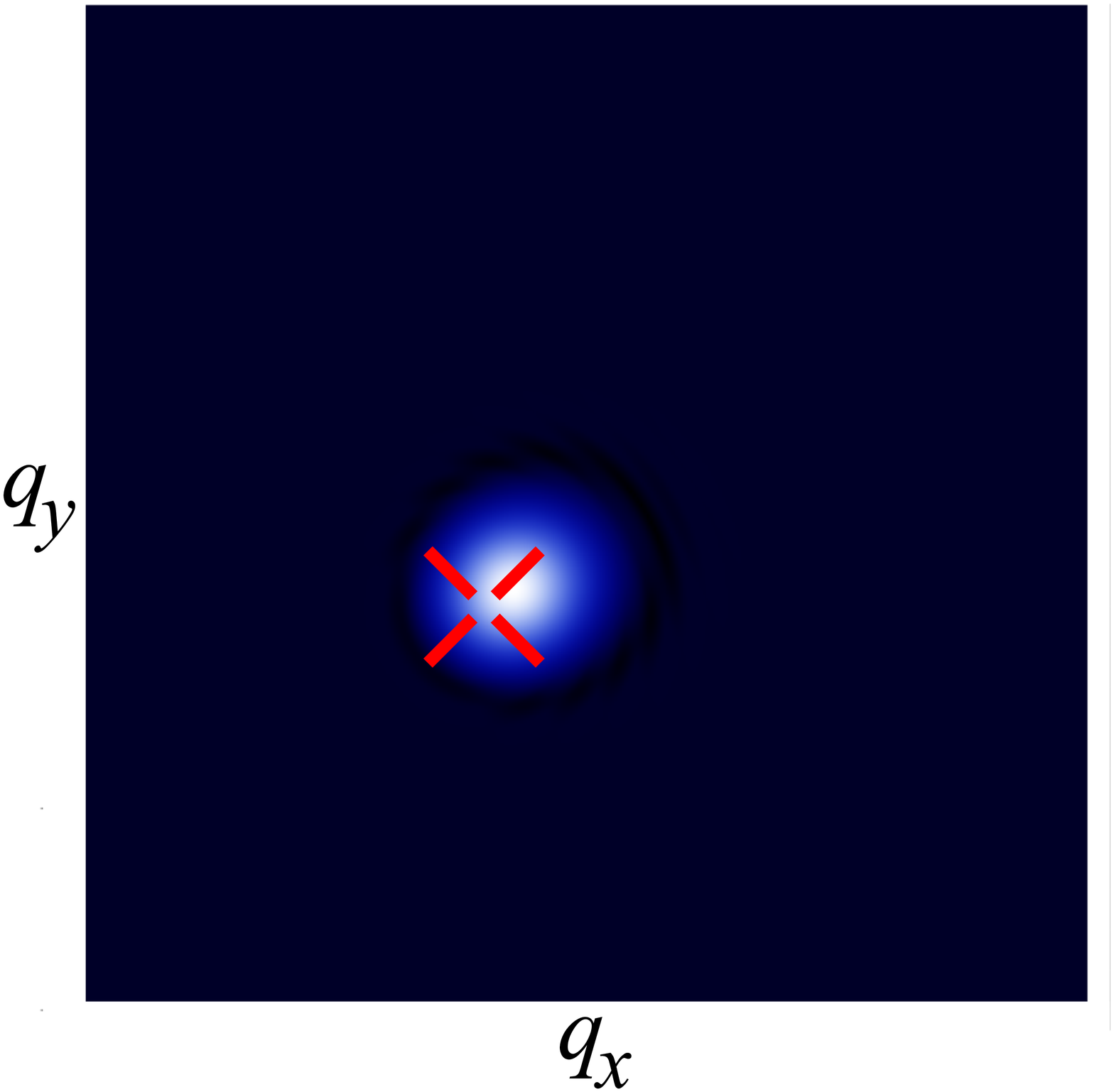}}
\caption{\label{damping-map}Examples of the damping rate map in momentum space for (a) $\mu\neq0$ and (b) $\mu=0$ (with and without a Fermi surface of Weyl electrons respectively). Brightness depicts the rate $\gamma({\bf q})$ of spin wave damping, and the red crosshair shows the reference $\Delta{\bf Q}$ for the local wavevector ${\bf q}=0$. These are $q_z=0$ slices through the full 3D map. Observing patterns of this kind in the full Brillouin zone scan will indicate the Weyl-electron origin of damping and reveal the complete set of $\Delta{\bf Q}={\bf Q}_m-{\bf Q}_n$ wavevectors from which the individual node wavevectors ${\bf Q}_m$ can be deduced (assuming, for example, $\sum_m {\bf Q}_m=0$). The bright outer ring, which shrinks and closes when $2|\mu|<|\Omega|$, originates in the inter-band electron scattering and gains strength from the rapidly growing Weyl electron density of states. Note that various details in these maps, such as the anisotropy and ring sizes, will generally depend on the concrete spin-wave dispersion $\Omega({\bf q}+\Delta{\bf Q})$, polarization, type and orientation of magnetic order, as well as the chiralities and symmetries of the Weyl nodes.}
\end{figure}

\begin{figure}[!t]
\includegraphics[width=2.6in]{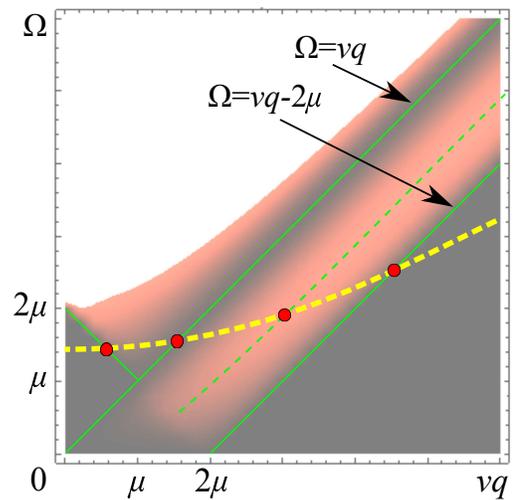}
\caption{\label{damp}A density plot of the collective mode damping rate $\gamma({\bf q},\Omega)$ induced by Weyl electrons. Thin solid green lines indicate $\gamma=0$, and the thin dashed green line indicates the local maximum of $\gamma$. The thick dashed yellow line represents the dispersion $\Omega({\bf q}+\Delta{\bf Q})$ of a hypothetical spin-wave excitation (note that the origin of the plot corresponds to the momentum difference $\Delta{\bf Q}$ of two Weyl nodes in the first Brillouin zone). The spin-wave's damping rate will exhibit local minimums and maximums at the shown red points, which are characteristic for the relativistic spectrum of Weyl electrons. Resolving two of these points is enough for the determination of the Weyl Fermi velocity $v$ and the chemical potential $\mu$ of the Weyl nodes addressed via $\Delta{\bf Q}$. Resolving three points allows an independent verification that Weyl nodes are indeed responsible for the damping. The two-parameter scaling of the damping rate (\ref{DampingRate2}) across a range of energies is the most general signature of Weyl electrons, and can be used to verify the Weyl-electron origin of damping even if the visible spin wave dispersion does not cross any of the shown characteristic points.}
\end{figure}

The essential utility of the universal damping comes from its qualitative features that reflect the relativistic nature of Weyl electrons. If the Fermi energy $\mu$ lies away from the energy of the Weyl nodes, Fermi surfaces will form. Then, the spin wave damping rate is expected to exhibit a set of minimums and maximums as a function of the frequency $\Omega$ and momentum ${\bf q}$. The locations of these extremums depend on the parameters that characterize the Weyl nodes: Fermi velocity $v$, chemical potential $\mu$ and even their relative chiralities $\chi_m \chi_n = \pm 1$. Fig.\ref{damping-map} demonstrates how the locations ${\bf Q}_m$ of Weyl nodes can be extracted from the full Brillouin zone map of the spin wave's damping rate $\gamma({\bf q})$. Once the wavevectors ${\bf Q}_m$ are known, Fig.\ref{damp} illustrates how the observation of enough extremums enables indirect measurements of the Weyl electron spectra on multiple Weyl nodes. The presence of Weyl Fermi pockets also introduces spin-momentum locking into the damping rates ($\gamma_{mn}^{\perp\perp'}\neq 0$), but only on the pairs of Weyl nodes with opposite chiralities. As a consequence, the two spin wave modes that carry opposite spin currents at the same wavevector ${\bf q}$ have different peak widths in inelastic neutron scattering.

The above qualitative features of damping disappear if the Fermi energy sits exactly at the Weyl nodes. However, the damping rate then becomes a universal function of a single parameter $|\Omega|/vq$. This kind of scaling is a signature of the relativistic Weyl electrons -- it is caused by ``inter-band'' transitions in which an electron below the Weyl node is excited to a state above the Weyl node. The plots of universal functions $f_{mn}^{ab}$ that appear in Eq. \ref{DampingRate2} at $\mu=0$ are shown in Fig.\ref{damping-all}.

The magnitude of the damping rate depends on the Kondo/Hund scale $J_{\textrm{K}}$ which may not be known. However, the spin wave damping caused by Weyl electrons is always related to the effective strength $J$ of the Weyl-electron-induced Ruderman–Kittel–Kasuya–Yosida (RKKY) interactions among the local moments \cite{Nikolic2020a}:
\begin{equation}
\frac{\gamma}{J} \sim \frac{1}{(a\Lambda)^{3}}\left(\frac{q}{\Lambda}\right)^{2}\times\left(\frac{\Omega}{vq}\right)^{2}
  \quad,\quad
J\sim v\Lambda\left(\frac{a^{3}\Lambda^{2}J_{\textrm{K}}}{v}\right)^{2} \ .
\end{equation}
Here, $\Lambda$ is the momentum cut-off for the linear Weyl spectrum, $|{\bf q}|<\Lambda$. Since $a\Lambda<1$ and the characteristic features of the universal damping appear near $|\Omega|\sim vq$, the damping rates are generally comparable to the energy scale $J$ of the induced RKKY interactions. For example, the RKKY energy scale in the magnetic Weyl semimetal NdAlSi\cite{Gaudet2021} can be crudely estimated as $J\sim 1\textrm{ meV}$. Even if the damping rate is more than an order of magnitude below this value of $J$, it should be detectable with high resolution neutron instruments (a spin echo spectrometer can achieve energy resolution below $10\;\mu\textrm{eV}$).

\raggedbottom

\section{Dissipation by Weyl electrons}\label{secDissipation}

Here we calculate the Gaussian dissipative part of the effective action for local moments which arises due to their coupling to itinerant Weyl electrons. The non-dissipative part of this action, computed in Ref.\cite{Nikolic2020a}, captures the induced RKKY interactions among the local moments: Heisenberg, Kitaev and Dzyaloshinskii-Moriya. All Gaussian terms $\delta n^a \Gamma^{ab} \delta n^b$ of the action obtain from a single two-point Feynman diagram which involves momentum integration of a singular function; the principal part of this integral yields the interactions, and the contribution of its pole singularity amounts to dissipation. We will focus only on the latter, following the procedure from Ref.\cite{Nikolic2020a}.

\begin{figure}
\includegraphics[width=1.3in]{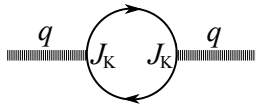}
\caption{\label{Bubble}The Feynman diagram for two-spin interactions. Thick external lines represent local moment fields and thin lines represent Weyl electron propagators. The two-spin couplings include Heisenberg, Kitaev and Dzyaloshinskii-Moriya interactions, but the Weyl-electron origin of spin dynamics also creates a dissipation channel in which spin waves decay into electron-hole pairs.}
\end{figure}

The essential dynamics of local moments $\hat{\bf n}_i$ coupled to conduction electrons $\psi_i$ is given by the Hamiltonian:
\begin{equation}\label{sdModel}
H_{0}^{\phantom{x}}=H_{n}^{\phantom{x}}+\sum_{{\bf k}}\epsilon_{{\bf k}}^{\phantom{x}}\psi_{{\bf k}}^{\dagger}\psi_{{\bf k}}^{\phantom{x}}+J_{\textrm{K}}^{\phantom{x}}\sum_i \hat{\bf n}_i^{\phantom{x}}\,\psi_i^{\dagger}\boldsymbol{\sigma}\psi_i^{\phantom{x}} \ .
\end{equation}
Both the local moments and electrons live on a lattice whose sites are labeled by $i$, but we will immediately take the continuum limit. The basic two-spin correlations $\langle \hat{n}_i^a \hat{n}_j^b \rangle$ are contained in the second-order Feynman diagram shown in Fig.\ref{Bubble}:
\begin{equation}\label{Gamma2b}
\Gamma^{ab}_{mn}(q) = i\frac{J_{\textrm{K}}^{2}}{2}\!\int\!\frac{d^{4}k}{(2\pi)^{4}} \textrm{tr}\left\lbrack G_{m}\!\left(k\!-\!\frac{q}{2}\right)\sigma^{a}G_{n}\!\left(k\!+\!\frac{q}{2}\right)\sigma^{b}\right\rbrack
\end{equation}
The Weyl electron Green's functions
\begin{equation}\label{Green2}
G_{n}(\omega,{\bf k})=\ \Bigl\lbrack \omega-H_{n}({\bf k})+i\,\textrm{sign}(\epsilon_{n}({\bf k}))0^{+} \Bigr\rbrack^{-1}
\vspace{0.1in}
\end{equation}
are treated as spinor matrices and refer to the low-energy electronic states near any Weyl node $n$ whose wavevector in the first Brillouin zone is ${\bf Q}_n$; the wavevector ${\bf k}$ is a ``small'' displacement $|{\bf k}|<\Lambda$ from ${\bf Q}_n$, where $\Lambda$ is the momentum cut-off for the linear Weyl dispersion. These low-energy electrons are described by the Hamiltonian
\begin{equation}\label{WeylH}
H_{n}({\bf k}+{\bf Q}_n)=v\chi_{n}\boldsymbol{\sigma}{\bf k}-\mu \ ,
\end{equation}
where $\mu$ is the chemical potential that determines the Weyl Fermi pocket character and size, $v$ is the Fermi velocity, and $\chi_n=\pm 1$ is the Weyl node chirality. We assume for simplicity that all Weyl nodes are spherically symmetric, share the same node energy, chemical potential and Fermi velocity, but have arbitrary wavevectors ${\bf Q}_n$ and chiralities $\chi_n=\pm 1$ (as long as the chiralities of all nodes in the first Brillouin zone add up to zero). By this construction, the expression (\ref{Gamma2b}) is associated with a pair $m,n$ of Weyl nodes, and ${\bf q}$ is a ``small'' wavevector measured relative to ${\bf Q}_m-{\bf Q}_n$.

We will carry out all calculations with the formal assumption that no external or effective magnetic field is exerted on electrons. Realistically, however, we are interested in magnetic Weyl semimetals whose local moments may carry a non-zero net magnetization $\hat{\bf n}_0$ that presents itself as an effective magnetic field ${\bf B} = -J_{\textrm{K}} \hat{\bf n}_0$ to electrons. This is of no concern because the correction of the spectrum (\ref{WeylH}) amounts merely to a shift of the wavevector ${\bf k} \to {\bf k} - {\bf B}/v\chi_n$. Hence, an effective magnetic field only alters the locations ${\bf Q}_n$ of the Weyl nodes in momentum space, which are arbitrary in our formalism. 

The full effective action matrix $\Gamma$ for local moments takes contributions from all Weyl node pairs:
\begin{equation}\label{Gamma2c}
\Gamma({\bf Q},\Omega) = \sum_{m,n} \Gamma_{mn}({\bf Q}-{\bf Q}_m+{\bf Q}_n, \Omega) \ .
\end{equation}
In this sense, it is possible to experimentally address a particular pair of Weyl nodes, or a set of pairs, by probing the momentum space in the vicinity of ${\bf Q} \sim {\bf Q}_m-{\bf Q}_n$. The dissipative part of $\Gamma_{mn}$ will contain information about the addressed Weyl nodes.

\subsection{Calculation of the dissipative terms in the effective spin Lagrangian}\label{secEffLagrangian}

The calculation of (\ref{Gamma2b}) is lengthy, so we will only outline its key steps. The trace has been evaluated before \cite{Nikolic2020a}, and the frequency integration yields:
\begin{widetext}
\begin{eqnarray}\label{Gamma2d}
\Gamma^{ab}_{mn}(q) &=& -\frac{J_{\textrm{K}}^2}{2} \!\int\!\frac{d^{3}k}{(2\pi)^{3}}\Biggl\lbrack\frac{X^{ab}(\Omega,{\bf q};v\chi_m\left\vert {\bf k}\!-\!\frac{{\bf q}}{2}\right\vert+\frac{\Omega}{2} -\mu,{\bf k})}{2v\chi_m\left\vert {\bf k}\!-\!\frac{{\bf q}}{2}\right\vert}
\prod_{s=\pm1}\frac{\theta\left(\mu-v\chi_m\left\vert {\bf k}\!-\!\frac{{\bf q}}{2}\right\vert \right)}{\Omega+v\chi_m\left\vert {\bf k}\!-\!\frac{{\bf q}}{2}\right\vert-vs\chi_n\left\vert {\bf k}\!+\!\frac{{\bf q}}{2}\right\vert +i0^{+}F(s\chi_n,\chi_m)} \nonumber \\
&& \qquad\quad -\frac{X^{ab}(\Omega,{\bf q};-v\chi_m\left\vert {\bf k}\!-\!\frac{{\bf q}}{2}\right\vert+\frac{\Omega}{2} -\mu,{\bf k})}{2v\chi_m\left\vert {\bf k}\!-\!\frac{{\bf q}}{2}\right\vert } \prod_{s=\pm1}\frac{\theta\left(\mu+v\chi_m\left\vert {\bf k}\!-\!\frac{{\bf q}}{2}\right\vert \right)}{\Omega-v\chi_m\left\vert {\bf k}\!-\!\frac{{\bf q}}{2}\right\vert -vs\chi_n\left\vert {\bf k}\!+\!\frac{{\bf q}}{2}\right\vert +i0^{+}F(s\chi_n,-\chi_m)} \nonumber \\
&& \qquad\quad +\frac{X^{ab}(\Omega,{\bf q};v\chi_n\left\vert {\bf k}\!+\!\frac{{\bf q}}{2}\right\vert -\frac{\Omega}{2}-\mu,{\bf k})}{2v\chi_n\left\vert {\bf k}+\frac{{\bf q}}{2}\right\vert } \prod_{s=\pm1}\frac{\theta\left(\mu-v\chi_n\left\vert {\bf k}\!+\!\frac{{\bf q}}{2}\right\vert \right)}{v\chi_n\left\vert {\bf k}\!+\!\frac{{\bf q}}{2}\right\vert -\Omega-vs\chi_m\left\vert {\bf k}\!-\!\frac{{\bf q}}{2}\right\vert -i0^{+}F(\chi_n,s\chi_m)} \nonumber \\
&& \qquad\quad -\frac{X^{ab}(\Omega,{\bf q};-v\chi_n\left\vert {\bf k}\!+\!\frac{{\bf q}}{2}\right\vert -\frac{\Omega}{2}-\mu,{\bf k})}{2v\chi_n\left\vert {\bf k}\!+\!\frac{{\bf q}}{2}\right\vert } \prod_{s=\pm1}\frac{\theta\left(\mu+v\chi_n\left\vert {\bf k}\!+\!\frac{{\bf q}}{2}\right\vert \right)}{-v\chi_n\left\vert {\bf k}\!+\!\frac{{\bf q}}{2}\right\vert -\Omega-vs\chi_m\left\vert {\bf k}\!-\!\frac{{\bf q}}{2}\right\vert -i0^{+}F(-\chi_n,s\chi_m)} \nonumber
\end{eqnarray}
\end{widetext}
Here, $\theta(x)$ is the step function, and two more functions, $X^{ab}({\Omega,{\bf q};\omega,{\bf k}})$ and $F(s_+,s_-)$ are introduced to simplify notation. The function $X^{ab}({\Omega,{\bf q};\omega,{\bf k}})$ obtains from the numerator of the trace in (\ref{Gamma2b}). Introducing the Kronecker symbol $\delta^{ab}$ and the Levi-Civita symbol $\epsilon^{abc}$, we have:
\begin{eqnarray}\label{Xa}
&& X^{ab}\left(\Omega,{\bf q};\omega,{\bf k}\right) = \left\lbrack (\omega+\mu)^{2}-\frac{\Omega^{2}}{4}\right\rbrack \delta^{ab} \nonumber \\
&& \quad +v^{2}\chi_m\chi_n\left\lbrack 2\left(k^{a}k^{b}-\frac{q^{a}q^{b}}{4}\right)-\delta^{ab}\left(k^{c}k^{c}-\frac{q^{c}q^{c}}{4}\right)\right\rbrack \nonumber \\
&& \quad +iv\epsilon^{abc}\biggl\lbrack \chi_m\left(\omega+\frac{\Omega}{2}+\mu\right)\left(k^{c}-\frac{q^{c}}{2}\right) \nonumber \\
&& \qquad\qquad -\chi_n\left(\omega-\frac{\Omega}{2}+\mu\right)\left(k^{c}+\frac{q^{c}}{2}\right)\biggr\rbrack \ .
\end{eqnarray}
The function $F(s_+,s_-)$ with $s_+,s_-=\pm 1$ keeps track of the infinitesimal imaginary terms in the denominators of Green's functions:
\begin{eqnarray}
&& F(s_+,s_-) = \textrm{sign}\left(vs_+\left\vert{\bf k}\!+\!\frac{\bf q}{2}\right\vert-\mu\right) - \textrm{sign}\left(vs_-\left\vert{\bf k}\!-\!\frac{\bf q}{2}\right\vert\right) \nonumber \\
&& \qquad = \theta\!\left(|{\bf qk}|-\left\vert \left(\frac{\mu}{v}\right)^{2}-k^{2}-\frac{q^{2}}{4}\right\vert \right) \nonumber \\
&& \qquad\qquad \times\left\lbrack \left(\textrm{sign}(\mu)+\frac{s_{+}+s_{-}}{2}\right)\textrm{sign}({\bf qk})+\frac{s_{+}-s_{-}}{2}\right\rbrack \nonumber \\
&& \qquad\quad + (s_{+}-s_{-})\,\theta\!\left(k^{2}+\frac{q^{2}}{4}-|{\bf q}{\bf k}|-\left(\frac{\mu}{v}\right)^{2}\right) \ .
\end{eqnarray}
At this point, we use the relationship
\begin{equation}
\frac{1}{x\pm i0^+} = \mathbb{P}\frac{1}{x} \mp i\pi\delta(x)
\end{equation}
to isolate the dissipative processes that curb the $x\to 0$ resonances. Dropping all terms that involve the principal part $\mathbb{P}$, we get:
\begin{eqnarray}\label{Gamma2e}
&& \widetilde{\Gamma}^{ab}_{mn}(q) = \frac{i\pi J_{\textrm{K}}^{2}}{8v^2} \sum_{s_m,s_n} \! s_ms_n \!\int\!\frac{d^{3}k}{(2\pi)^{3}}\frac{F'(s_n\chi_n,s_m\chi_m)}{\chi_m\chi_n\left\vert {\bf k}-\frac{{\bf q}}{2}\right\vert \left\vert {\bf k}+\frac{{\bf q}}{2}\right\vert } \nonumber \\
&& \quad \times X^{ab}\left(\Omega,{\bf q};vs_m\chi_m\left\vert {\bf k}-\frac{{\bf q}}{2}\right\vert +\frac{\Omega}{2}-\mu,{\bf k}\right) \\
&& \quad \times \;\delta\left(\Omega+vs_m\chi_m\left\vert {\bf k}-\frac{{\bf q}}{2}\right\vert -vs_n\chi_n\left\vert {\bf k}+\frac{{\bf q}}{2}\right\vert \right) \nonumber \\
&& \quad \times \left\lbrack \theta\left(\mu-vs_m\chi_m\left\vert {\bf k}-\frac{{\bf q}}{2}\right\vert \right)-\theta\left(\mu-vs_n\chi_n\left\vert {\bf k}+\frac{{\bf q}}{2}\right\vert \right)\right\rbrack \nonumber
\end{eqnarray}
We introduced $F'=\textrm{sign}(F)\,(1-\delta_{F,0})$, and the sum goes over $s_m, s_n = \pm 1$. All chirality factors $\chi_m, \chi_n = \pm 1$ that appear outside of $X^{ab}$ are clearly eliminated by the summation over $s_m, s_n$, so it will be convenient do define $s_- = s_m\chi_m = \pm1$ and $s_+ = s_n\chi_n = \pm1$. The Dirac $\delta$-function in (\ref{Gamma2e}) imposes:
\begin{equation}
s_{+}\left\vert {\bf k}+\frac{{\bf q}}{2}\right\vert -s_{-}\left\vert {\bf k}-\frac{{\bf q}}{2}\right\vert =\frac{\Omega}{v} \ .
\end{equation}
This pins the magnitude of the wavevector ${\bf k}$ to
\begin{equation}\label{Kval}
k=\frac{|\Omega|}{2v}\sqrt{\frac{\Omega^{2}-v^{2}q^{2}}{\Omega^{2}-v^{2}q^{2}\cos^{2}\theta}} \ ,
\end{equation}
assuming ${\bf q}{\bf k} = qk\cos\theta$, and further requires satisfying one of these two conditions:
\begin{equation}
\begin{array}{lcl}
|\Omega|>vq & \quad\wedge\quad & s_{\pm}=\pm\textrm{sign}(\Omega) \\
|\Omega|<vq|\cos\theta| & \quad\wedge\quad & s_{+}=s_{-}=\textrm{sign}(\Omega\cos\theta)
\end{array} \nonumber
\end{equation}\newline
The wavevector ${\bf k}=(k,\theta,\phi)$ integration in (\ref{Gamma2e}) is now conveniently performed in the spherical coordinate system referenced to the external wavevector ${\bf q}$. The integral over $k=|{\bf k}|$ is immediately solved due to the Dirac $\delta$-function and we merely need to replace the occurrences of $|{\bf k}|$ with (\ref{Kval}). The integral over $\phi$ affects only the quantities (\ref{Xa}) leading to $\int d\phi X^{ab} = 2\pi v^2 \widetilde{X}^{ab}$ with the following non-zero components:
\begin{eqnarray}\label{Xb}
\widetilde{X}^{\parallel\parallel} &=& s_{+}s_{-}\left\vert {\bf k}-\frac{{\bf q}}{2}\right\vert \left\vert {\bf k}+\frac{{\bf q}}{2}\right\vert \\
&& +\chi_{m}\chi_{n}\left\lbrack k^{2}(2\cos^{2}\theta-1)-\frac{q^{2}}{4}\right\rbrack \nonumber \\
\widetilde{X}^{\perp\perp} &=& s_{+}s_{-}\left\vert {\bf k}-\frac{{\bf q}}{2}\right\vert \left\vert {\bf k}+\frac{{\bf q}}{2}\right\vert \nonumber \\
&& +\chi_{m}\chi_{n}\left(-k^{2}\cos^{2}\theta+\frac{q^{2}}{4}\right) \nonumber \\
\widetilde{X}^{\perp\perp'} &=& i\epsilon^{\parallel\perp\perp'} \biggl\lbrack -\left(\chi_{m}s_{+}\left\vert {\bf k}+\frac{{\bf q}}{2}\right\vert +\chi_{n}s_{-}\left\vert {\bf k}-\frac{{\bf q}}{2}\right\vert \right)\frac{q}{2} \nonumber \\
&& + \left(\chi_{m}s_{+}\left\vert {\bf k}+\frac{{\bf q}}{2}\right\vert -\chi_{n}s_{-}\left\vert {\bf k}-\frac{{\bf q}}{2}\right\vert \right)k\cos\theta \biggr\rbrack \nonumber \ .
\end{eqnarray}
Here and onward, the upper spin indices denote directions $\parallel$ parallel to ${\bf q}$, and two mutually perpendicular directions $\perp,\perp'$ which are also perpendicular to ${\bf q}$. Note that $\epsilon^{\parallel\perp\perp'}$ implements a chiral ``right-hand-rule'' relationship between the three spin directions. The integral over $\theta$ is finite and conveniently evaluated numerically. At the end, we arrive at:
\begin{equation}\label{Gamma2f}
\widetilde{\Gamma}^{ab}_{mn}(q) = i\,\frac{J_{\textrm{K}}^{2} \Omega^2}{128\pi v^3}\; f_{mn}^{ab}\left(\frac{|\Omega|}{vq},\frac{|\Omega|}{2|\mu|};\textrm{sign}(\mu,\Omega)\right)
\end{equation}
where the dimensionless functions $f = \alpha + \beta$ have contributions from intra-band $\alpha$ and inter-band $\beta$ electron scattering. Note that the inter-band processes require transferring an electron between the two states whose energies have opposite signs, and thus can occur only when $|\Omega|>2|\mu|$. Defining
\begin{equation}
\lambda = \frac{vq}{|\Omega|} \quad,\quad x = \frac{2|\mu|}{|\Omega|} \quad,\quad
\kappa = \sqrt{\frac{1-\lambda^2}{1-\lambda^2\xi^2}}
\end{equation}
with $|\xi|=|\cos\theta|$, we have: 
\begin{widetext}
\begin{eqnarray}\label{GammaDetails1}
\alpha_{mn}^{\perp\perp} &=& \int\limits_{0}^{1}d\xi\,\theta\left(2\kappa\lambda\xi-|x^2-\kappa^2-\lambda^2|\right)\,\kappa^{2} \\
&& \times\left\lbrack \left(1-\chi_{m}\chi_{n}\frac{-\kappa^{2}\xi^{2}+\lambda^{2}}{\sqrt{(\kappa^{2}+\lambda^{2})^{2}-(2\kappa\lambda\xi)^{2}}}\right)\theta(1-\lambda)+\left(1+\chi_{m}\chi_{n}\frac{-\kappa^{2}\xi^{2}+\lambda^{2}}{\sqrt{(\kappa^{2}+\lambda^{2})^{2}-(2\kappa\lambda\xi)^{2}}}\right)\theta(\lambda\xi-1)\right\rbrack \nonumber \\[0.1in]
\alpha_{mn}^{\parallel\parallel} &=& \int\limits_{0}^{1}d\xi\,\theta\left(2\kappa\lambda\xi-|x^2-\kappa^2-\lambda^2|\right)\,\kappa^{2} \nonumber \\
&& \times\left\lbrack \left(1-\chi_{m}\chi_{n}\frac{\kappa^{2}(2\xi^{2}-1)-\lambda^2}{\sqrt{(\kappa^{2}+\lambda^{2})^{2}-(2\kappa\lambda\xi)^{2}}}\right)\theta(1-\lambda)+\left(1+\chi_{m}\chi_{n}\frac{\kappa^{2}(2\xi^{2}-1)-\lambda^2}{\sqrt{(\kappa^{2}+\lambda^{2})^{2}-(2\kappa\lambda\xi)^{2}}}\right)\theta(\lambda\xi-1)\right\rbrack \nonumber \\[0.1in]
\alpha_{mn}^{\perp\perp'} &=& -i\,\epsilon^{\parallel\perp\perp'} \int\limits_{0}^{1}d\xi\,\theta\left(2\kappa\lambda\xi-|x^2-\kappa^2-\lambda^2|\right)\,\kappa^{2} \nonumber \\
&& \times\sum_{s=\pm1} \frac{(\chi_{m}+\chi_{n})\, \textrm{sign}(\mu)+s\,(\chi_{m}-\chi_{n})\, \textrm{sign}(\Omega)}{2\sqrt{\kappa^2+\lambda^2-2s\kappa\lambda\xi}}\, (\kappa\xi-s\lambda)\, \Bigl\lbrack \theta(1-\lambda)-s\,\theta(\lambda\xi-1)\Bigr\rbrack \nonumber \ ,
\end{eqnarray}
\end{widetext}
and
\begin{widetext}
\begin{eqnarray}\label{GammaDetails2}
\beta_{mn}^{\perp\perp} &=& \int\limits_{-1}^{1}d\xi\,\theta\left(\kappa^2+\lambda^2-2\kappa\lambda|\xi|-x^2\right)\,\kappa^{2} \left(1-\chi_{m}\chi_{n}\frac{-\kappa^2\xi^{2}+\lambda^2}{\sqrt{\left(\kappa^2+\lambda^2\right)^{2}-(2\kappa\lambda\xi)^2}}\right)\,\theta(1-\lambda) \\
\beta_{mn}^{\parallel\parallel} &=& \int\limits_{-1}^{1}d\xi\,\theta\left(\kappa^2+\lambda^2-2\kappa\lambda|\xi|-x^2\right)\,\kappa^{2} \left(1-\chi_{m}\chi_{n}\frac{\kappa^2(2\xi^{2}-1)-\lambda^2}{\sqrt{\left(\kappa^2+\lambda^2\right)^{2}-(2\kappa\lambda\xi)^2}}\right)\,\theta(1-\lambda) \nonumber \\
\beta_{mn}^{\perp\perp'} &=& -i\,\epsilon^{\parallel\perp\perp'} \int\limits_{-1}^{1}d\xi\,\theta\left(\kappa^2+\lambda^2-2\kappa\lambda|\xi|-x^2\right)\,\kappa^{2} \sum_{s=\pm1} \frac{s\,(\chi_{m}-\chi_{n})\, \textrm{sign}(\Omega)}{2\sqrt{\kappa^2+\lambda^2-2s\kappa\lambda|\xi|}}\,(\kappa|\xi|-s\lambda)\,\theta(1-\lambda) \nonumber \ .
\end{eqnarray}
\end{widetext}

The functions $f^{ab}$ have the same characteristics in all spin channels $a,b\in\lbrace \perp\perp, \parallel\parallel, \perp\perp'\rbrace$. Their plots in Figures \ref{damping-tl}, \ref{damping-all}, \ref{damp} illustrate that $f^{ab}$ vanish for $|\Omega|<v|{\bf q}|-2|\mu|$, $2|\mu|-v|{\bf q}|>|\Omega|>v|{\bf q}|$ and $|\Omega|=v|{\bf q}|$. The dissipation at $|\Omega|>\textrm{max}(2|\mu|,v|{\bf q})$ is dominated by the collective mode decay into ``high energy'' particle-hole pairs which are excited across the Weyl node. Outside of this frequency-momentum region, the decay occurs by generating ``low energy'' particle-hole pairs across the Fermi surface on the Weyl node. This ``low energy'' channel is weaker, but has several features that clearly reveal the relativistic properties of the Weyl spectrum. Fig. \ref{damp} shows how the minimums and maximums of a collective mode damping rate can be used to characterize the Fermi surface of Weyl electrons.

\subsection{Spin wave damping}\label{secSpinWaves}

The actual damping rate of collective excitations generally obtains from a mixture of spin channels. Consider the spin waves with wavevectors $\Delta{\bf Q} + {\bf q}$ in the vicinity of the momentum-space separation $\Delta{\bf Q} = {\bf Q}_m - {\bf Q}_n$ between two particular Weyl nodes. Let $-S\Omega_0^{ab}({\bf q})$ be the intrinsic part of the effective Lagrangian density $\delta\mathcal{L}_{\textrm{eff}}$ for the local moment fluctuations $\delta{\bf n}$, excluding the spin Berry phase $S\Omega\delta^{ab}$ ($S$ is the spin magnitude of local moments). This can contain any exchange interactions of the localized electrons and crystal field anisotropies. The Lagrangian density terms induced by the itinerant Weyl electrons are all contained in the $\Gamma^{ab}$ tensor (\ref{Gamma2b}). The principal part of (\ref{Gamma2b}) yields a variety of induced RKKY interactions \cite{Nikolic2020a}, while its dissipative components $\widetilde{\Gamma}^{ab}$ are collected in (\ref{Gamma2f}). The presence of magnetic order in the ground state further affects the dynamics of spin waves because the small spin fluctuations $\delta{\bf n}$ of low-energy modes must be orthogonal to the local spins $\hat{\bf n}$. This can be incorporated into the general analysis \cite{Nikolic2019b}, but we will simplify the discussion here by considering only a ferromagnetic ground state $\hat{\bf n}({\bf r}) = \hat{\bf n}_0$. The spectrum of damped spin waves is extracted from the Gaussian part of the Lagrangian density in momentum space
\begin{equation}\label{Leff}
\delta\mathcal{L}_{\textrm{eff}} = (\delta n^a)^{*} \Bigl\lbrack S\Omega\delta^{ab} - S\Omega_0^{ab}({\bf q}) + a^3\,\Gamma^{ab}(\Omega,{\bf q}) \Bigr\rbrack \delta n^b
\end{equation}
The factor of a unit-cell volume $a^3$ converts the energy density $\Gamma^{ab}$ to the energy per lattice unit-cell, and the factor of $\frac{1}{2}$ in the Berry phase term $\Omega$ is appropriate for the local moments with spin $S=\frac{1}{2}$. Introducing
\begin{equation}
g^{ab} = \Omega_0^{ab} - \frac{a^3}{S}\, \Gamma^{ab}
\end{equation}
to simplify notation, the spin wave modes obtain by diagonalizing $\mathcal{P}\mathcal{M}\mathcal{P}$, where $\mathcal{P}^{ab}=\delta^{ab}-\hat{n}_{0}^{a}\hat{n}_{0}^{b}$ projects-out the high-energy amplitude fluctuations (keeps $\delta{\bf n} \perp \hat{\bf n}_0$) and
\begin{equation}
\mathcal{M}^{ab} = \Omega\delta^{ab}-g^{\perp\perp}(\delta^{ab}-\hat{q}^{a}\hat{q}^{b})
  -g^{\parallel\parallel}\hat{q}^{a}\hat{q}^{b}-g^{\perp\perp'}\epsilon^{abc}\hat{q}^{c}
  \nonumber
\end{equation}
is the matrix embedded in (\ref{Leff}). An arbitrary choice of the background magnetization $\hat{\bf n}_0 = \hat{\bf z}$ reveals two polarization modes $\delta{\bf n}=(\delta n^{x},\delta n^{y})$ at ${\bf q} = q \hat{\bf q}$
\begin{equation}
\delta{\bf n}_{\pm} \propto \left(\begin{array}{c}
  \frac{g^{\parallel\parallel}-g^{\perp\perp}}{2}(\hat{q}_{x}^{2}-\hat{q}_{y}^{2})
    \pm \delta\epsilon \\
  (g^{\parallel\parallel}-g^{\perp\perp})\hat{q}_{x}\hat{q}_{y}-g^{\perp\perp'}\hat{q}_{z}
\end{array}\right)
\end{equation}
with energies
\begin{equation}\label{SWdisp}
\Omega_{\pm} = g_{0}^{\perp\perp}+
  \frac{g^{\parallel\parallel}-g^{\perp\perp}}{2}(1-\hat{q}_{z}^{2}) \pm \delta\epsilon
\end{equation}
where $\delta\epsilon = \frac{1}{2}\sqrt{(g^{\parallel\parallel}-g^{\perp\perp})^{2}(1-\hat{q}_{z}^{2})^{2}-(2g^{\perp\perp'}\hat{q}_{z})^2}$. These polarizations are generally elliptical, but become circular $\delta{\bf n} \propto (\pm i, 1)$ with $\Omega_{\pm} = g^{\perp\perp} \mp ig^{\perp\perp'}$ for the modes that propagate along the magnetization direction (${\bf q} \parallel \hat{\bf n}_0$), and linear $\delta{\bf n}_+ \propto \hat{\bf q}$, $\delta{\bf n}_- \propto \hat{\bf n}_0\times\hat{\bf q}$ with $\Omega_+ = g^{\parallel\parallel}$, $\Omega_- = g^{\perp\perp}$ respectively for the modes that propagate in the plane perpendicular to the magnetization (${\bf q}\perp\hat{\bf n}_0$). The character and non-degeneracy of the two polarization modes is the hallmark of the RKKY interactions induced through the spin-orbit coupling: Dzyaloshinskii-Moriya (DM) in the case of circular polarizations, and Kitaev in the case of linear polarizations.

The equation (\ref{SWdisp}) has to be solved self-consistently since the components of the $g^{ab}$ tensor on its right-hand side depend on frequency, but the revealed form of its solutions ensures all of the spin wave properties that we discuss. The two circular polarizations at the same wavevector ${\bf q} \parallel \hat{\bf n}_0$ carry opposite spin currents
\begin{equation}
j_{i}^{a} = -iq_{i}\epsilon^{abc}(\delta n^{b})^{*}\delta n^{c} \propto \mp |g^{\perp\perp'}|^{2}q_{i}\delta^{az} \ ,
\end{equation}
so their energy difference $\Omega_{\pm} = g^{\perp\perp} \mp ig^{\perp\perp'}$ due to the DM interaction implies spin-momentum locking. Note that the DM interactions appears as $g^{ab}_{\textrm{DM}} \propto \epsilon^{abc}(i q^c)$, so it does shift the spin wave energy. The dissipative components $\widetilde{g}^{ab} \propto \widetilde{\Gamma}^{ab}$ of $g^{ab}$ impart an imaginary part on the pole frequency $\Omega$, which corresponds to the damping rate. The signs of both $\widetilde{\Gamma}^{\perp\perp}, \widetilde{\Gamma}^{\parallel\parallel}$ ($f^{\perp\perp}, f^{\parallel\parallel} > 0$) indeed correspond to damping and not an instability, and the chiral contributions are not large enough to overturn this at any $\Omega$. The chiral dissipative part extracted from (\ref{Gamma2f}) is real, $\widetilde{g}^{ab}_{\textrm{DM}} \propto \epsilon^{abc}q^c$, and hence introduces different damping rates for the two circular spin waves. These qualitative conclusions hold for the elliptical modes as well.

\subsection{The absence of uniform precession damping}\label{secFieldEq}

The universal dependence of (\ref{Gamma2f}) on $|\Omega|/vq$ introduces a non-analytic behavior at $\Omega,q\to0$ in the damping terms $\widetilde{\mathcal{L}}$ of the spin Lagrangian density. Therefore, one cannot strictly expand $\widetilde{\mathcal{L}}$ in powers of $\Omega,q$ to represent the dissipation as a result of local processes. $\widetilde{\mathcal{L}}$ can be approximated by an expansion only in special limits. Suppose the spin waves have dispersion $|\Omega|=uq$ at low energies (in the vicinity of $\Delta{\bf Q}={\bf Q}_m-{\bf Q}_n \to 0$ for intra-node scattering $m=n$). If the spin wave velocity $u$ is smaller than the Weyl electrons' velocity $v$, then a sufficiently large $q$ pushes the spin waves into the regime $|\Omega|< vq-2|\mu|$ where $\widetilde{\Gamma}^{ab} = 0$ in (\ref{Gamma2f}) and the damping is absent (see Fig.\ref{damping-all}). Alternatively, if $u\gg v$, then the spin waves are in the regime $|\Omega|\gg vq$ and their damping at energies $|\Omega|>2|\mu|$ is approximately characterized by the dominant local terms $\widetilde{\Gamma}^{\parallel\parallel}, \widetilde{\Gamma}^{\perp\perp} \sim i(A\Omega^{2}+Bq^{2})$ and a smaller chiral term $\widetilde{\Gamma}^{\perp\perp'} \sim Dq\Omega$. Together with the non-dissipative Hermitian terms $\chi_0^{-1}$, the electron-induced part of the local moments' effective Lagrangian density (\ref{Leff}) contains
\begin{equation}\label{ChiT}
\Gamma^{ab} \xrightarrow{|\Omega|\gg vq} \frac{1}{2} \Bigl\lbrack(\chi_0^{-1})^{ab} 
  + i(A^{ab}\Omega^{2}\!+\!B^{ab}q^{2}) + D\epsilon^{abc}q^c\Omega \Bigr\rbrack
\end{equation}
with $A^{ab} = A^{\perp\perp}(\delta^{ab}-q^aq^b/q^2) + A^{\parallel\parallel}q^aq^b/q^2$ and likewise for $B^{ab}$. By construction (\ref{Gamma2b}), $\Gamma \equiv \frac{1}{2}\chi^{-1}$ is the inverse time-ordered correlation function
\begin{equation}
\langle \delta s^{a}({\bf q},\Omega)\, \delta s^{b}({\bf q}',\Omega')\rangle
  = i \chi^{ab}({\bf q},\Omega) \,\delta({\bf q}+{\bf q}')\delta(\Omega+\Omega') \nonumber
\end{equation}
for the small fluctuations $\delta{\bf s}$ of the Weyl electron spins away from their equilibrium magnetization. We will consider only the simplest case of a collinear ferromagnet in the following analysis. The equilibrium state will be given by the uniform magnetization of local moments $\hat{\bf n}_0$ and electrons $\langle{\bf s}_0\rangle \parallel \hat{\bf n}_0$. 

A semiclassical representation of the local moment dynamics is given by the field equation for $\hat{\bf n}$. The presence of non-Hermitian damping terms in the effective action for local moments prevents us from deriving the field equation by considering the stationary action condition. Instead, we can use linear response theory to learn about the semiclassical dynamics. The retarded electrons' spin correlation function
\begin{equation}
\chi_{\textrm{R}}({\bf q},\Omega) =
  \begin{cases}
    \chi({\bf q},\Omega) & ,\quad\Omega>0\\
    \chi^{\dagger}({\bf q},\Omega) & ,\quad\Omega<0
  \end{cases}
\end{equation}
is readily obtained from (\ref{ChiT})
\begin{eqnarray}\label{ChiR}
&& (\chi_{\textrm{R}}^{-1})^{ab} \xrightarrow{|\Omega|\gg vq} (\chi_{0}^{-1})^{ab} \\
&& \qquad\qquad + \textrm{sign}(\Omega) \Bigl\lbrack 
  i(A^{ab}\Omega^{2}\!+\!B^{ab}q^{2}) + D\epsilon^{abc}q^c\Omega \Bigr\rbrack \ , \nonumber
\end{eqnarray}
and then the response of electron spins to the local moment field is
\begin{equation}
\langle \delta s^{a}({\bf r},t) \rangle = \frac{J_{\textrm{K}}}{a^3} \int dt'd^{3}r'\,
  \chi_{\textrm{R}}^{ab}({\bf r}-{\bf r}',t-t')\,\delta n^{b}({\bf r}',t') \ .
\end{equation}
This follows from the Kondo interaction $J_{\textrm{K}}$ in (\ref{sdModel}) between the ``perturbation'' field ${\bf n}$ and the responding electrons spin ${\bf s} = \psi^\dagger \boldsymbol{\sigma} \psi$ on a lattice site (the unit-cell volume $a^3$ effectively converts the integration over coordinates to a summation over lattice sites). Note that $\chi_{\textrm{R}}^{ab}({\bf q},\Omega) = (\chi_{\textrm{R}}^{ab})^*(-{\bf q},-\Omega)$ is established globally in momentum space (not necessarily in the immediate vicinity of the Weyl node wavevector $\Delta{\bf Q}$) \footnote{The chiral damping term in Eq.\ref{ChiR} naively violates this property, but one must take into account $D\propto \chi_m-\chi_n$, which stems from (\ref{GammaDetails1}) and (\ref{GammaDetails2}), to recover the needed property globally in momentum space (an electron scattering from Weyl node $m$ to node $n$ in the vicinity of $\Delta{\bf Q}={\bf Q}_m-{\bf Q}_n$ is accompanied by the opposite scattering from node $n$ to $m$ in the vicinity of $-\Delta{\bf Q}$).}, so that its inverse Fourier transform $\chi_{\textrm{R}}^{ab}({\bf r},t)$ is real. The thermodynamic potential for local moments is simply
\begin{equation}
F\lbrack\hat{{\bf n}}\rbrack = J_{\textrm{K}}\langle{\bf s}\rangle\hat{{\bf n}} \ .
\end{equation}
The local moment dynamics is driven by an effective ``magnetic'' field in units of energy
\begin{equation}
{\bf H}_{\textrm{eff}}({\bf r},t)=-\frac{\delta F[\hat{\bf n}]}{\delta\hat{\bf n}({\bf r},t)}=-J_{\textrm{K}}\langle{\bf s}({\bf r},t)\rangle
\end{equation}
Taking into account the Berry phase of local moments yields the usual semiclassical field equation
\begin{equation}\label{Feq1}
\frac{\partial\hat{\bf n}}{\partial t} = \hat{\bf n}\times{\bf H}_{\textrm{eff}} \ .
\end{equation}
with
\begin{eqnarray}\label{Feq2}
H_{\textrm{eff}}^{a}({\bf r},t) &\approx& -J_{\textrm{K}} \hat{n}_0^a
    -\frac{J_{\textrm{K}}^2}{a^3}\int d\delta td^{3}\delta r\,
       \chi_{\textrm{R}}^{ab}(\delta{\bf r},\delta t) \\
&& \qquad\quad \times \,\delta\hat{n}^{b}({\bf r}+\delta{\bf r},t+\delta t) \nonumber \\
&\approx& -J_{\textrm{K}} \hat{n}_0^a -\frac{J_{\textrm{K}}^2}{a^3}\int d\delta td^{3}\delta r\,
    \chi_{\textrm{R}}^{ab}(\delta{\bf r},\delta t) \nonumber \\
&& \qquad\quad \times
  \left\lbrack 1+\delta{\bf r}\boldsymbol{\nabla}+
  \delta t\frac{\partial}{\partial t}+\cdots\right\rbrack \delta\hat{n}^{b}({\bf r},t)
  \nonumber
\end{eqnarray}
This is seen to generate Gilbert damping which dissipates the precession of uniform magnetization in typical ferromagnets
\begin{equation}
\frac{\partial\hat{\bf n}}{\partial t}=\hat{\bf n}\times{\bf H}_{\textrm{eff}}=\cdots
  + \hat{\bf n} \times \alpha_{\textrm{G}} \frac{\partial\hat{\bf n}}{\partial t}
\end{equation}
with the damping tensor
\begin{eqnarray}
\alpha_{\textrm{G}}^{ab} &=& -\frac{J_{\textrm{K}}^2}{a^3}\int d\delta td^{3}\delta r\,\chi_{\textrm{R}}^{ab}(\delta{\bf r},\delta t)\,\delta t \\
 &=& i\,\frac{J_{\textrm{K}}^2}{a^3}\int\frac{d\Omega}{2\pi}\int d\delta t\,e^{-i\Omega\delta t}\,
     \frac{\partial\chi_{\textrm{R}}^{ab}(0,\Omega)}{\partial\Omega} \nonumber \\
 &=& -\frac{J_{\textrm{K}}^2}{a^3}\frac{\partial\,\textrm{Im}\,\chi_{\textrm{R}}^{ab}}{\partial\Omega}
     \biggr\vert_{({\bf q},\Omega)=0} \ . \nonumber
\end{eqnarray}
The real part of $\chi_{\textrm{R}}({\bf q},\Omega)$ generally does not contribute because it is an even function of $\Omega$ at $q=0$ (even though it diverges for gapless spin waves when $\Omega\to 0$). In the case of damping induced by Weyl electrons, the imaginary part of $\chi_{\textrm{R}}$ becomes zero when $2|\mu|-vq>|\Omega|\ge vq$, following the behavior of the time-ordered $\chi^{-1} \equiv \Gamma^{ab}$ that was discussed earlier (see Fig.\ref{damp}). Therefore, $\chi_{\textrm{R}}$ is real in the limit $\Omega,q\to 0$ and the decay of spin waves into Stoner excitations of the Weyl electrons does not generate Gilbert damping.

The complete equation of motion for local moments can be extracted from (\ref{Feq1}) and (\ref{Feq2}), but the non-analytic frequency dependence of the dissipative terms in (\ref{ChiR}) introduces (via its Fourier transform) non-local relationships between the fields $\hat{\bf n}(t)$ at different times $t$. If one were to ignore this issue, or approximate the non-local effect by couplings over small time intervals, then a local field equation would be obtained from the expansion indicated in (\ref{Feq2}). We will not pursue this here any further.

\section{Conclusions and discussion}\label{secConclusions}

We analyzed the dynamics of local magnetic moments coupled to itinerant Weyl electrons, and focused on the dissipation of spin waves via the continuum of Stoner particle-hole excitations.  We described this dissipation at the level of the effective Lagrangian of local moments, or equivalently the spin-spin correlation function (dynamic susceptibility). For the spin waves at wavevector $\Delta{\bf Q}+{\bf q}$ and frequency $\Omega$ in the vicinity of the momentum difference $\Delta{\bf Q}={\bf Q}_m-{\bf Q_n}$ between two Weyl nodes, the damping rate is proportional to $\Omega^2$ and a universal function of $|\Omega|/v|\bf q|$ where $v$ is the Weyl electron (Fermi) velocity. The presence of Fermi pockets with chemical potential $\mu$ introduces additional dependence of the damping rate on $|\Omega/\mu|$. If the Weyl nodes are well-separated in momentum space, then there is no cross-talk between them in the damping rates and the momentum-space locations of the Weyl nodes can be discerned from the wavevectors at which the spin wave dissipation is locally maximized. The Weyl-electron origin of dissipation can be experimentally verified by the universal relativistic properties of damping over a range of mode frequencies and momenta, while various parameters of the Weyl spectrum can be extracted from the momentum space locations of the characteristic damping features (e.g. local maximums and points where damping vanishes). The damping rates involving Weyl electrons also generally exhibit ``non-reciprocity'' or chirality -- the modes of different polarizations that propagate at the same momentum ${\bf q}$ have different lifetimes. We presented a procedure to obtain the field equation for the semi-classical dynamics of the local moment magnetization field, and found that the dissipation on Weyl electrons does not give rise to Gilbert damping.

One important conclusion of this study is that the spin wave damping rate reveals the relativistic nature of Weyl electrons -- both through its universal dependence on $|\Omega|/v|\bf q|$ and the places in momentum space where it vanishes. We computed the damping rate associated with Stoner excitations, but similar results should hold for zero-spin particle-hole excitations as well. Then, other kinds of collective modes coupled to Weyl electrons, e.g. the phonons of the crystal or a charge density wave, should exhibit similar universality in their damping rates. This would be interesting to explore in the future since inelastic neutron scattering is sensitive to phonons as well.

The developed theory is very general within its limitations. It makes no assumptions about the Weyl node locations, so it applies to Dirac semimetals as well (where the opposite-chirality Weyl nodes coexist at the same wavevectors). It also makes no assumptions about the magnetic order, so it holds for ferromagnets, antiferromagnets and paramagnets, with or without local spin anisotropy. In this regard, however, the damping rates of spin waves are affected by the nature of magnetic order; we demonstrated the calculations only in the ferromagnetic (and implicitly also the paramagnetic) case. Analytical progress was made by simplifying the model to spherically symmetric Weyl nodes that all live at the same energy. This is the main limitation of the current theory, although many implications of realistic model extensions can be readily anticipated. Energy differences between the nodes are easily included by associating different chemical potentials to the nodes, while a small Weyl node anisotropy is expected to introduce a similar anisotropy in the induced dynamics and dissipation of local moments. It is possible that type-II Weyl nodes fall outside of this theory's domain, so their exploration is left for future study. We also did not consider corrections due to finite temperature and disorder.

The usefulness of this theory for the experimental characterization of magnetic Weyl semimetals is guarantied in principle, but depends on several factors in reality. The needed level of detail is not easy to achieve in the measurements of spin wave spectra. It requires at least very clean samples, low temperatures, as well as a sufficiently high energy resolution and adequate statistics to resolve with low error bars the energy/momentum dependence of the inelastic neutron scattering. These aspects of measurements can always be improved, but there are also material-related constraints: phonons, for example, must not coexist with spin waves at the same momenta and frequencies. Still, some regions of the first Brillouin zone should expose the electronic damping mechanism and enable the proposed experimental characterization of magnetic Weyl semimetals. On the purely theoretical front, the present study was concerned with a basic but intricate and important aspect of interaction physics in a topological system. It plays a role in piecing together a broader picture of magnetic correlated topological materials, which can host non-trivial anisotropic magnetic orders \cite{Gaudet2021}, chiral magnetic states and excitations \cite{Nikolic2019b}, and possibly even exotic spin liquids \cite{Nikolic2019}.

\section{Acknowledgments}\label{secAck}

I am very grateful for insightfull discussions with Jonathan Gaudet and Collin Broholm. This research was supported at the Institute for Quantum Matter, an Energy Frontier Research Center funded by the U.S. Department of Energy, Office of Science, Basic Energy Sciences under Award No. DE-SC0019331.

\appendix


%

\end{document}